\documentclass[iop,twocolappendix,
twocolumn]{aastex701}
\usepackage{amsmath}
\usepackage{amssymb}
\usepackage{bm}
\usepackage{graphicx}
\usepackage{color}
\usepackage{float}
\usepackage{multirow}
\usepackage{CJK}
\usepackage[T1]{fontenc}
\usepackage[hang,flushmargin]{footmisc}

\hypersetup{
  colorlinks,
  linkcolor={blue!88!black!80},
  citecolor={blue!88!black!80},
  urlcolor={blue!88!black!80}}

\definecolor{darkgreen}{rgb}{0,0.35,0}
\definecolor{blue}{rgb}{0,0,1}
\definecolor{red}{rgb}{0,0,0} 

\newcommand{\be}{\begin{eqnarray}}
\newcommand{\ee}{\end{eqnarray}}

\newcommand\yps{\bgroup\markoverwith{\textcolor[rgb]{1.0, 0.0, 1.0}{\rule[0.5ex]{8pt}{1.5pt}}}\ULon}

\begin{document}
\begin{CJK*}{UTF8}{gbsn} 

\shorttitle{}
\title{Multi-wavelength Constraints on Dust Dynamics and Size Evolution in Protoplanetary Disk Rings. II.  Observational Implications}

\author[0009-0002-8049-6554]{Linhan Yang (杨林翰)}
\affiliation{Shanghai Astronomical Observatory, Chinese Academy of Sciences, Shanghai 200030, People's Republic of China}
\affiliation{School of Astronomy and Space Science, University of Chinese Academy of Sciences, Beijing 100049, People's Republic of China}
\email[]{}

\author[0000-0002-7329-9344]{Ya-Ping Li (李亚平)}\thanks{liyp@shao.ac.cn}
\affiliation{Shanghai Astronomical Observatory, Chinese Academy of Sciences, Shanghai 200030, People's Republic of China}
\email[]{liyp@shao.ac.cn}

\author[0000-0001-9290-7846]{Ruobing Dong (董若冰)}
\affiliation{Kavli Institute for Astronomy and Astrophysics, Peking University, Beijing 100871, People's Republic of China}
\email[]{}

\author[0000-0003-3728-8231]{Yinhao Wu (吴寅昊)}
\thanks{EACOA Fellow}
\affiliation{Shanghai Astronomical Observatory, Chinese Academy of Sciences, Shanghai 200030, People's Republic of China}
\affiliation{National Astronomical Observatory of Japan, 2-21-1 Osawa, Mitaka, Tokyo 181-8588, Japan}
\email[]{}

\author[0000-0003-2300-2626]{Hauyu Baobab Liu (\begin{CJK*}{UTF8}{ipxm}呂浩宇\end{CJK*})}
\affiliation{Department of Physics, National Sun Yat-Sen University, No. 70, Lien-Hai Road, Kaohsiung City 80424, Taiwan, ROC}
\affiliation{Center of Astronomy and Gravitation, National Taiwan Normal University, Taipei 116, Taiwan, ROC}
\email[]{}

\author[0000-0003-1958-6673]{Kiyoaki Doi (\begin{CJK*}{UTF8}{ipxm}土井聖明\end{CJK*})}
\affiliation{Max-Planck Institute for Astronomy, Königstuhl 17, D-69117 Heidelberg, Germany}
\email[]{}

\author[0000-0002-5991-8073]{Anibal Sierra}
\affiliation{Universidad Nacional Aut\'{o}noma de M\'{e}xico. Instituto de Astronom\'{i}a. A.P. 70-264, 04510. Ciudad de M\'{e}xico, M\'{e}xico}
\email[]{}

\author[0000-0002-7002-8928]{Greta Guidi}
\affiliation{University Grenoble Alpes, CNRS, IPAG, 38000 Grenoble, France}
\email[]{}

\author[0000-0002-7575-3176]{Pinghui Huang (黄平辉)}
\affiliation{CAS Key Laboratory of Planetary Sciences, Purple Mountain Observatory, Chinese Academy of Sciences, Nanjing 210008, People's Republic of China}
\email[]{}

\begin{abstract}
Spatially resolved dust rings in protoplanetary disks are widely used to infer disk and dust physics from multi-wavelength continuum observations.
Their interpretation, however, often neglects grain growth and the evolution of the size distribution, limiting the connection between observed ring profiles and dust-evolution parameters.
Building on a physical dust-ring model that includes coagulation and fragmentation, we develop a Bayesian inference framework that jointly incorporates radiative transfer and finite angular resolution.
When applied to two rings in HD~163296 and two in LkCa~15, our framework yields gas-dependent estimates of the key dust-evolution parameters such as turbulence strength $\alpha$ and the fragmentation velocity $v_{\rm frag}$ in a self-consistent way.
\textcolor{red}{Most rings admit both a low-$\alpha$, low-$v_{\rm frag}$ branch with small grains, and a higher-$\alpha$, higher-$v_{\rm frag}$ branch with larger grains.}
\textcolor{red}{Typical low-$\alpha$ branches have $\alpha\sim10^{-5}$--$10^{-4}$ and fragmentation velocities of order cm~s$^{-1}$ level, whereas the higher-$\alpha$ branches reach $\alpha\sim10^{-3}$--$10^{-2}$ and fragmentation velocities of a few to $20~{\rm m~s^{-1}}$.}
\textcolor{red}{The observed broad and wavelength-dependent profiles near the ring peaks can be reproduced by intrinsically narrow dust rings.}
This new framework offers a more direct route from multi-wavelength continuum data to the microphysics of dust growth and trapping---a connection that can be robustly tested with future high-resolution observations at longer wavelengths.

\end{abstract}

\keywords{Protoplanetary disks (1300); Dust continuum emission (412); Bayesian statistics (1900)}

\section{Introduction}
\label{sec:intro}

High-resolution Atacama Large Millimeter/submillimeter Array (ALMA) observations have revealed that ring-like substructures are ubiquitous in protoplanetary disks \citep[e.g.,][]{ALMA2015,Andrews2018,Long2018,Sierra2021,Bae2023}. These dust rings are commonly interpreted as a direct tracer of dust trapping \textcolor{red}{(i.e., the radial drift of dust grain toward a local pressure maximum through aerodynamic drag between dust and gas; e.g., } \citealt{Pinilla2012,Long2020,Sierra_2025}). \textcolor{red}{Other scenarios cannot be fully ruled out,} like dust-induced instabilities \citep{Gonzalez2017}, traffic jam effects \citep{Jiang_2021,Carrera2021} and snowlines \citep{Zhang2015,Okuzumi2016}.
The leading mechanism for the dust trapping is the density perturbations induced by embedded planets \citep[e.g.,][]{Dong_2015a,Dong_2018, Zhang_2018,Li2019b,Paardekooper2023,Wu2023HD163296}. It has been widely used to infer disk properties and dust physics \citep[e.g.,][]{Doi_2023,Liu2024}.

However recovering the underlying physical parameters from multi-wavelength ring observations is challenging because the mapping from dust physics to observed continuum profiles involves three layers of complexity:
\begin{enumerate}
    \item The underlying dust growth and transport within a pressure bump, where coagulation, fragmentation, drift, and diffusion jointly shape the radial dust distribution.
    \item The radiative-transfer mapping from dust structure to continuum emission, which depends on wavelength-dependent opacity, optical depth, absorption, and scattering.
    \item The modification of the emergent profile by finite angular resolution and noise in real interferometric observations.
\end{enumerate}
Interpreting multi-wavelength ring widths therefore requires a framework that connects these three layers in a self-consistent way.

Previous studies have not yet treated these three layers in a single, systematic inference framework, often neglecting dust growth and fragmentation in the underlying dust distribution \citep{Dullemond_2018}, or treating the observed continuum brightness profile as a direct tracer of that distribution without self-consistently modeling radiative transfer and beam convolution \citep{Huang_2018,Long2018,rosotti_2020}.

Moreover, observed ring width-wavelength trends are diverse, ranging from narrower rings at longer wavelengths to nearly wavelength-independent or non-monotonic behavior \citep{Macias_2019,Sierra2021,Tazzari-etal.2021,Karina_2021,Carvalho_2024,Shi_2026}. This diversity suggests that no single simple mechanism can explain all multi-wavelength ring profiles, further motivating a self-consistent forward-modeling approach.

\citet{Yang_2025} addressed the first of these complexities by developing a dust-trapping model that includes growth and fragmentation. \textcolor{red}{They showed that growth--fragmentation equilibrium can produce nearly size-independent radial dust distributions within rings,} in sharp contrast to the commonly adopted drift--diffusion-only picture in which smaller grains are expected to form broader rings and larger grains remain more narrowly trapped \citep{Birnstiel_2010,Dullemond_2018,rosotti_2020}. This result provides a more physically consistent baseline for interpreting multi-wavelength rings beyond the drift--diffusion-only picture.

Here we address the remaining two layers of complexity by combining the dust-ring model of \citet{Yang_2025} with radiative transfer, finite angular resolution, and Bayesian inference in a self-consistent framework implemented in the open-source Python package \texttt{dring}\footnote{\url{https://github.com/SScoorge/dring}.} \citep{yang_2026_soft}. This framework enables us to connect physical ring parameters to observable multi-wavelength continuum profiles and to reassess what can, and cannot, be robustly inferred from multi-wavelength rings. 

\section{Methods}\label{sec:app_method}

Our method combines the dust-ring model of \citet{Yang_2025}, radiative transfer with finite angular resolution, and Bayesian inference to connect physical ring parameters to observed multi-wavelength continuum profiles. We present our physical dust ring model in Section~\ref{subsec:dustringmodel}, intensity profile of the dust ring in Section~\ref{subsec:RT_Res}, and Bayesian inference of model parameters in Section~\ref{subsec:Bayesian}.

\subsection{Physical Dust-Ring Model} 
\label{subsec:dustringmodel}

Following \citet{Yang_2025}, we adopt the standard dust-trapping picture in which a localized gas-pressure maximum concentrates radially drifting solids into a ring. The pressure bump is prescribed as a Gaussian function,
\begin{equation}
P(r)=P_0\exp\!\left[-\frac{1}{2}\left(\frac{r-r_0}{w_{\rm p}}\right)^2\right],
\end{equation}
where \textcolor{red}{$r$ is the disk radial coordinate from the central star,} $r_0$ and $w_{\rm p}$ denote the gas bump center and width, and $P_0\equiv P(r_0)=P_0(\Sigma_{\rm g,0},T_0)$ is set by the gas surface density normalization $\Sigma_{\rm g,0}$ and the midplane temperature $T_0$. Here and below, the subscript ``0'' denotes quantities evaluated at $r_0$. Gas and dust are assumed to share the same temperature profile, $T\propto T_0\,r^{-0.5}$. We focus on the local region around the pressure maximum, where the continuum ring is expected to form and dominate the emission.

Dust growth and transport within the bump are modeled following Section~5.3 of \citet{Yang_2025}. \textcolor{red}{In that framework, dust evolves in an axisymmetric, prescribed gas pressure bump. The dust collision velocities include contributions from Brownian motion, turbulence, radial and azimuthal drift, and vertical settling, and the collision outcomes include sticking/growth, fragmentation, and erosion, while bouncing is not included. Dust is radially transported by drift and turbulent diffusion. Because the gas pressure profile is prescribed and fixed in time, gas advection associated with gas evolution and dust feedback on the gas are not included.} \textcolor{red}{For the analytic description used below,} the grain-size number distribution at $r_0$ is assumed to follow a power law,
\begin{equation}
n(a)\propto a^{-q},
\end{equation}
and throughout this work we adopt $q=3.5$. \textcolor{red}{The impact of varying $q$ is discussed in Appendix~\ref{appendix:Robustness}.} The maximum dust size is set by the fragmentation barrier $a_{\rm frag}$ \citep{Birnstiel_2010}, obtained by equating the relative collision velocity $v_{\rm rel}(a,\alpha,\Sigma_{\rm g},T,\rho_{\rm s})$ \citep[e.g., Figure. 7 in][]{Birnstiel_2011} 
to the fragmentation threshold $v_{\rm frag}$, where $\alpha$ is the turbulent strength parameter \citep{ShakuraSunyaev} and $\rho_{\rm s}$ is the bulk density of dust grains.

Most importantly, \citet{Yang_2025} showed that, inside a gas pressure bump, the 1D radial dust-to-gas ratio profile of each dust species can be approximated by a Gaussian function centered at $r_0$, with size-dependent standard deviation $w_{\epsilon,a}$ in the growth-fragmentation equilibrium. Here $w_{\epsilon,a}$ denotes the intrinsic Gaussian standard width of the dust-to-gas ratio profile for grains of size $a$.
For grains with $a<0.37\,a_{\rm frag,0}$, the Gaussian width is nearly size-independent, and we denote this constant width by  $w_{\epsilon}$,

\begin{equation}
w_\epsilon
\simeq
1.2w_{\rm p}\sqrt{\frac{5-q}{4-q}\frac{\alpha}{\mathrm{St}_{\max}}},
\end{equation}
Here $\mathrm{St}_{\max}\equiv \mathrm{St}_0(a_{\rm frag,0})$ is the Stokes number evaluated at $r_0$ for the fragmentation-limited maximum size.

By contrast, grains with $a\gtrsim 0.37\,a_{\rm frag,0}$ remain confined to a narrower region around $r_0$, and their width can be approximated by $w_{\epsilon,a}\propto \rm{St}_0^{-0.5}(a)$. The numerical factor $0.37$ should be understood as an empirical transition size calibrated in \citet{Yang_2025}, rather than as a sharp physical boundary. 


Physically, this transition can be understood in the following way. For $a \lesssim 0.37\,a_{\rm frag,0}$, continual growth and fragmentation exchange material among grain sizes fast enough that different dust species share nearly the same radial distribution. By contrast, grains with $a \gtrsim 0.37\,a_{\rm frag,0}$ are close to the maximum size allowed at the pressure maximum. Their drift timescales are shorter, and because the local fragmentation barrier decreases away from the bump center, such grains can survive only in the central part of the ring. They therefore become more radially confined than the smaller grains. 

We impose two physical boundaries on the parameter space where this analytic description is applied. First, we require $\mathrm{St}_{\max}<1$, because particles with unity Stokes number can grow too efficiently and rapidly deplete the dust reservoir, in tension with the observed continuum rings. Second, the fitting formulae above require a moderately high total dust-to-gas ratio at the pressure maximum (e.g., with a typical dust-to-gas ratio of $\epsilon_{0} \gtrsim 10^{-2}$, see Appendix A.1 of \citet{Yang_2025} for details) for frequent collisions to maintain the growth-fragmentation-regulated dust distribution. 
Below these thresholds, dust collisions become too infrequent for the above approximation to remain valid, and drift or diffusion can more directly reshape the dust distribution. \citet{Yang_2025} provided fitting formulae for such low-collision-frequency regimes in its appendix, but in this work we would treat solutions in these regions as low-confidence because attaining such a low $\epsilon$ regime is generally unrealistic.
None of posterior solutions in this work falls into this ultra-low-$\epsilon$ regime.

\begin{figure*}[t!]
\centering
\includegraphics[width=\textwidth]{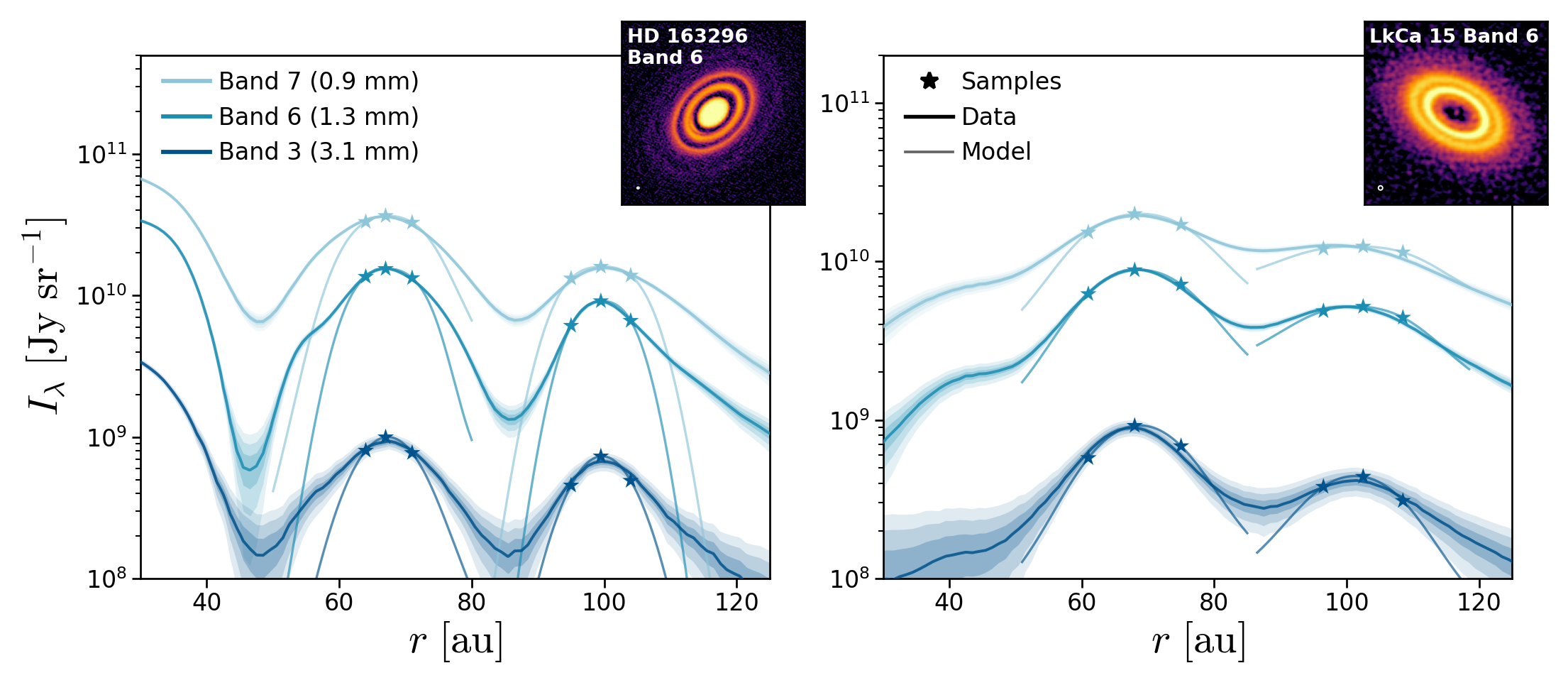}
\caption{
Radial intensity profiles of the dust continuum emission for the two sources (HD~163296 in left panel and LkCa~15 in right panel) 
analyzed in this work. 
The disk images at Band 6 for the two disks are shown in the upper right corner of each panel.
Colored curves show the observed radial profiles at the wavelengths listed in the legends \textcolor{red}{(the listed wavelengths are approximate reference values)}, with shaded regions indicating the observational uncertainties adopted in the likelihood calculation (corresponding to the $1\sigma$, $2\sigma$, and $3\sigma$ confidence levels).
Thin colored curves show the best-fit model profiles.
Stars mark the model intensities sampled at the radii used in the likelihood evaluation.
}
\label{fig:infit}
\end{figure*}

\subsection{From Dust Structure to Continuum Profiles} 
\label{subsec:RT_Res}

For a given dust-ring model, the unconvolved radial continuum intensity profile $I_{\rm raw}(r)$ is computed by including both absorption and scattering opacity using the analytic prescription of \citet{zhu_2019}, with disk inclination accounted for through the projected optical depth. Unless otherwise noted, we adopt the default \texttt{DSHARP} opacity \citep{Birnstiel_2018}.
\textcolor{red}{For each wavelength, we define the absorption and effective scattering optical depths as}
\begin{align}
\textcolor{red}{\tau_{\rm abs}(r)} &\textcolor{red}{= \int \kappa_{\rm abs}(a)\Sigma_{\rm d}(r,a)\,{\rm d}\ln a,}\\
\textcolor{red}{\tau_{\rm sca,eff}(r)} &\textcolor{red}{= \int (1-g)\kappa_{\rm sca}(a)\Sigma_{\rm d}(r,a)\,{\rm d}\ln a,}
\end{align}
\textcolor{red}{The scattering asymmetry parameter, $g=\langle \cos\theta \rangle$, is the intensity-weighted average of the cosine of the scattering angle over all scattering directions and is computed from the Mie-theory scattering phase function for the adopted grain properties and wavelength \citep{Birnstiel_2018}. It ranges from $-1$ to $1$, with $g=-1$, $0$, and $1$ corresponding to purely backward, isotropic, and purely forward scattering, respectively. The total optical depth and effective albedo are}
\begin{equation}
\textcolor{red}{
\tau_{\rm d}=\tau_{\rm abs}+\tau_{\rm sca,eff},
\qquad
\omega=\frac{\tau_{\rm sca,eff}}{\tau_{\rm d}} .
}
\end{equation}
\textcolor{red}{The emergent intensity at inclination $i$ is then}
{\color{red}
\begin{equation}
\begin{aligned}
I_{\rm raw}(r)
&=
B_\nu[T(r)]
\left(1-e^{-\tau_{\rm d}/\mu}\right)\\
&\times
\left[
1-\omega
\frac{
\begin{gathered}
e^{-\sqrt{3(1-\omega)}\tau}
+
e^{\sqrt{3(1-\omega)}(\tau-\tau_{\rm d})}
\end{gathered}
}{
\begin{gathered}
e^{-\sqrt{3(1-\omega)}\tau_{\rm d}}
\left(1-\sqrt{1-\omega}\right)
+
\left(1+\sqrt{1-\omega}\right)
\end{gathered}
}
\right],
\end{aligned}
\end{equation}
}
\textcolor{red}{where $\mu=\cos i$ and}
\begin{equation}
\textcolor{red}{
\tau=\frac{2\mu\tau_{\rm d}}{3\tau_{\rm d}+1}.
}
\end{equation}

Finite angular resolution is represented by convolving $I_{\rm raw}(r)$ with a one-dimensional Gaussian smoothing kernel,
\begin{equation}
I^{\rm conv}(r)
=
\int_{-\infty}^{+\infty}
I_{\rm raw}(r')\,G(r-r';\sigma_{\rm res})\,{\rm d}r' ,
\end{equation}
where
\begin{equation}
G(r-r';\sigma_{\rm res})
=
\exp\!\left[
-\frac{(r-r')^{2}}{2\sigma_{\rm res}^{2}}
\right]/\sqrt{2\pi}\sigma_{\rm res} .
\end{equation}
For convenience, we write the corresponding one-dimensional Gaussian width as $\sigma_{\rm res}=\theta_{\rm res}\cdot d/2.355$, where $d$ is the source distance and $\theta_{\rm res}$ denotes the effective angular resolution of the observation, i.e., the FWHM of the
effective Gaussian point-spread function. 

\subsection{Bayesian Inference of Physical Parameters} \label{subsec:Bayesian}






Our goal is to infer the physical parameters that control the multi-wavelength continuum profiles of dust rings within the framework described above. In this work, the parameters inferred by the Bayesian framework are the turbulence strength $\alpha$, the fragmentation velocity $v_{\rm frag}$, the mid-plane temperature at the pressure maximum $T_0$, and the dust-to-gas ratio at the pressure maximum $\epsilon_0$. The ring center $r_0$, the pressure-bump width $w_{\rm p}$, and the gas surface density normalization at the bump center $\Sigma_{\rm g,0}$ are held fixed using independent constraints \textcolor{red}{summarized in Appendix~\ref{app:hd163296_setup}--\ref{app:lkca15_setup}}, because continuum radial profiles alone provide limited leverage on these quantities. For a given parameter vector
\begin{equation}
\boldsymbol{\theta} = (\alpha,\, v_{\rm frag},\, T_0,\, \epsilon_0),
\end{equation}
the dust-ring model in Section~\ref{subsec:dustringmodel} and the radiative-transfer calculation in Section~\ref{subsec:RT_Res} together predict the continuum radial intensity profile at each observed wavelength, including the effect of finite angular resolution.


We constrain $\boldsymbol{\theta}$ by comparing the forward-modeled radial profiles with the observed multi-wavelength continuum profiles. At each wavelength $\lambda_n$, the model provides a convolved intensity profile $I^{\rm conv}_{\lambda_n}(r \mid \boldsymbol{\theta})$, which is evaluated at a set of sampled radii $\{r_m\}$. Because the absolute flux-calibration uncertainty is a band-wide multiplicative uncertainty rather than an independent random error at each radius, we introduce one multiplicative calibration factor $c_{\lambda_n}$ for each wavelength. These calibration factors are included in the Bayesian inference as nuisance parameters and are sampled together with the physical parameters $\boldsymbol{\theta}$.

In addition, we retain the native angular resolution of each wavelength, so some bands may be oversampled in radius. We account for this by correcting the effective chi-square contribution with a Nyquist factor,
\begin{equation}
N_{{\rm Nyquist},\lambda_n}
=
\max\left(
\frac{\theta_{{\rm res},\lambda_n} d}{\Delta r_{\lambda_n}},
1
\right),
\end{equation}
where $\Delta r_{\lambda_n}$ is the radial sampling interval of the profile at wavelength $\lambda_n$. When $\Delta r_{\lambda_n}<\theta_{{\rm res},\lambda_n}d$, the profile is oversampled and the corresponding chi-square is divided by $N_{{\rm Nyquist},\lambda_n}$; otherwise $N_{{\rm Nyquist},\lambda_n}=1$.

The likelihood is then
\begin{equation}
\begin{aligned}
\mathcal{L}(\boldsymbol{\theta},\boldsymbol{c})
\propto&
\prod_{n=1}^{N_{\lambda}}
\prod_{m=1}^{N_r}
\exp\left[
-\frac{1}{2}
\left(
\frac{
c_{\lambda_n} I^{\rm conv}_{\lambda_n}(r_m \mid \boldsymbol{\theta})
-
I^{\rm obs}_{\lambda_n}(r_m)
}{
\sigma_{{\rm rms},\lambda_n,m}\sqrt{N_{{\rm Nyquist},\lambda_n}}
}
\right)^2
\right] \\
&\times
\prod_{n=1}^{N_{\lambda}}
\exp\left[
-\frac{1}{2}
\left(
\frac{
c_{\lambda_n}-1
}{
f_{{\rm cal},\lambda_n}
}
\right)^2
\right] .
\end{aligned}\label{eq:likelihood}
\end{equation}
Here $\boldsymbol{c}=\{c_{\lambda_n}\}$ denotes the set of band-dependent calibration factors, $\sigma_{{\rm rms},\lambda_n,m}$ is the random measurement uncertainty of the radial profile, and $f_{{\rm cal},\lambda_n}$ is the adopted fractional absolute flux-calibration uncertainty for that band. This treatment preserves the correlated nature of the calibration uncertainty within each band, while the radial measurement errors are treated as independent Gaussian errors.

The posterior probability distribution is then given by Bayes' theorem,
\begin{equation}
P(\boldsymbol{\theta},\boldsymbol{c}\mid {\rm data})
\propto
\mathcal{L}(\boldsymbol{\theta},\boldsymbol{c})\,P(\boldsymbol{\theta},\boldsymbol{c}),
\end{equation}
where $P(\boldsymbol{\theta},\boldsymbol{c})$ denotes the prior as discussed in Appendix \ref{app_prior}. We sample the posterior using \texttt{UltraNest}\citep{ultranest1}.

The validity of the inference framework is assessed through end-to-end mock observations. 
We construct realistic multi-wavelength synthetic datasets and analyze them with the same profile-extraction and fitting procedure adopted for the real data. These tests verify that the simplified analytic framework used here can recover the main ring properties and the key parameter degeneracies relevant to this work. Throughout the analysis, we preserve the native angular resolution of each band, rather than homogenizing all datasets to a common beam. 
Full details of the mock setup and validation are deferred to Appendix~\ref{Appendix:Mock}.


\section{Results}\label{sec:results}
In this section, we apply the Bayesian inference framework to multi-wavelength ALMA ring profiles and constrain the physical parameters. We first present the fits to HD~163296 and LkCa~15 in Section~\ref{subsec:fit_results}, then discuss the posterior degeneracies revealed by the multi-wavelength data in Section~\ref{subsec:posterior_degeneracies}. We next interpret the resulting constraints on $\alpha$ and $v_{\rm frag}$ together with their dependence on the assumed gas properties in Section~\ref{subsec:alpha_vfrag_constraints}, and finally summarize the derived optical depths, dust masses, and intrinsic ring widths in Section~\ref{subsec:derived_quantities}.

\subsection{Multi-Wavelength Fits to HD~163296 and LkCa~15}
\label{subsec:fit_results}
We apply the framework to two prominent rings in each of the well-studied disks HD~163296 and LkCa~15. The details of the fitting setups are described in Appendix~\ref{appendix:fitting_setups}. In both systems, the inner and outer rings are hereafter referred to as Ring~1 and Ring~2, respectively. We perform separate fits for the two rings in each disk, as their physical conditions could differ.
As shown in Figure~\ref{fig:infit}, we use azimuthally averaged radial intensity profiles measured from the CLEAN images presented by \citet{Guidi_2022} and \citet{Sierra_2025}. 
For consistency, we restrict the analysis of both systems to Bands~3, 6, and 7. 
These targets are selected as representative high-quality multi-wavelength datasets, rather than as a survey; the method itself can be readily applied to other disks, such as MWC 480 \citep{Shi_2026}.
The observed profiles and fitting-configuration files used in this work are included in the public \texttt{dring} repository.

\textcolor{red}{The likelihood evaluation is restricted to the bright ring component around each pressure maximum, where collisions are frequent enough to maintain the growth--fragmentation-regulated dust distribution assumed in our analytic model. Far from the ring peak, the dust-to-gas ratio decreases rapidly and the dust physics can be more complex, so the analytic description may no longer be applicable. These regions are therefore outside the scope of the present local ring model. In practice, forcing the single-ring model to fit these outer regions can overfit emission outside the local ring peak and artificially reduce the physical degeneracies that should remain in the ring-only inference. At the same time, the fitted radial range cannot be made arbitrarily narrow because a region much smaller than the beam contains little independent radial information after convolution. Based on our tests, we therefore choose sampled radii that cover roughly one to two beam widths around each ring peak. Modest changes in the exact sampled radii mainly affect the posterior uncertainties, but do not change the dominant degeneracies or the order of magnitude of the inferred physical parameters. The adopted sampled radii $\{r_m\}$ are listed in Appendix~\ref{appendix:fitting_setups}.}

\textcolor{red}{Within the sampled radial ranges,} our framework provides satisfactory fits to the observed multi-wavelength radial ring profiles for both HD~163296 and LkCa~15 as shown in Figure~\ref{fig:infit}. 
For each ring of two disks, the best-fitting model simultaneously matches the observed profiles in intensity and ring width across multiple wavelengths.

\textcolor{red}{We note, however, that the model does not reproduce all low-level emission near the ring edges. This discrepancy is most evident at the inner edge of HD~163296 Ring~1, where the azimuthally averaged profile may be affected by crescent-like asymmetric emission in the disk. The other rings follow a similar trend, though to a weaker degree, with the observed emission remaining broader near the edges than the local ring model predicts. We therefore interpret the present fits as constraints on the bright ring component near the pressure maximum, rather than as a complete description of the surrounding gap and inter-ring regions. Understanding the dust distribution in these transition regions will require future work, likely involving full numerical dust-evolution simulations that can follow the coupled evolution of dust in both rings and adjacent gaps.}



\begin{deluxetable*}{lcccccccccc}
\tablecaption{
Multi-wavelength Bayesian fits to ALMA dust rings (Column (1)).
For each ring, we report posterior constraints on the turbulence strength $\alpha$ (2), fragmentation velocity $v_{\rm frag}$ (3), mid-plane temperature $T_0$ (4), and dust-to-gas ratio at the pressure maximum $\epsilon_0$ (5), together with the derived maximum grain size $a_{\max}$ (6), peak absorption optical depth at 1.3 mm $\tau_{0}^{\rm 1.3mm}$ (7), total dust mass $M_{\rm dust}$ (8), characteristic dust-to-gas ratio width $w_{\epsilon}$ (9), the average chi-square per fitted data point, $\chi^2_{\rm avg}\equiv\chi^2/N_{\rm data}$ (10), and the calibration factors $(c_7,c_6,c_3)$ (11). Here $c_7$, $c_6$, and $c_3$ correspond to ALMA Bands~7, 6, and 3, respectively.
We quote $\chi^2_{\rm avg}$ rather than the formal reduced chi-square because each ring is fitted with a small number of radial samples and the effective number of independent data points is affected by finite angular resolution.
Superscripts $A$, $B$, and $C$ in the first column denote distinct posterior modes where applicable. HD stands for HD 163296, and LkCa refers to LkCa 15.
Quoted uncertainties correspond to the 68\% credible intervals.
For parameters whose posteriors reach a prior boundary, we quote one-sided limits rather than median values.
}
\label{tab:ALMA_fits}
\tablehead{
\colhead{$\rm Ring^{mode}$} &
\colhead{$\alpha$} &
\colhead{$v_{\rm frag}$} &
\colhead{$T_0$} &
\colhead{$\epsilon_0$} &
\colhead{$a_{\max}$} &
\colhead{$\tau_{0}^{\rm 1.3mm}$} &
\colhead{$M_{\rm dust}$} &
\colhead{$w_{\epsilon}$} &
\colhead{$\chi^2_{\rm avg}$} &
\colhead{$(c_7,c_6,c_3)$} \\
\colhead{($r_0$ [au])} &
\colhead{$-$} &
\colhead{[cm s$^{-1}$]} &
\colhead{[K]} &
\colhead{$-$} &
\colhead{[cm]} &
\colhead{$-$} &
\colhead{[$M_{\oplus}$]} &
\colhead{[au]} &
\colhead{$-$} &
\colhead{$-$} 
}
\colnumbers
\startdata
$\mathrm{HD~}{\mathrm{Ring\ 1}}^{A}~(66.1)$ &
$5^{+2}_{-2}\times10^{-4}$ &
$214^{+83}_{-58}$ &
$24.2^{+1.6}_{-1.5}$ &
$2.1^{+0.4}_{-0.3}\times10^{-2}$ &
$1.7^{+0.7}_{-0.3}\times10^{-1}$ &
$0.8^{+0.1}_{-0.1}$ &
$68^{+9}_{-7}$ &
$4.4^{+0.4}_{-0.3}$ &
$0.9$ &
${\scriptstyle(1.1^{+0.1}_{-0.1},0.9^{+0.1}_{-0.1},1.0^{+0.1}_{-0.1})}$ \\
$\mathrm{HD~}{\mathrm{Ring\ 1}}^{B}~(66.1)$ &
$4^{+1}_{-1}\times10^{-5}$ &
$16^{+5}_{-4}$ &
$24.9^{+2.1}_{-2.0}$ &
$1.1^{+0.2}_{-0.1}\times10^{-1}$ &
$2.1^{+0.4}_{-0.3}\times10^{-2}$ &
$1.3^{+0.2}_{-0.2}$ &
$300^{+47}_{-36}$ &
$3.6^{+0.4}_{-0.3}$ &
$0.5$ &
${\scriptstyle(1.1^{+0.1}_{-0.1},0.9^{+0.1}_{-0.1},1.0^{+0.1}_{-0.1})}$ \\
$\mathrm{HD~}{\mathrm{Ring\ 1}}^{C}~(66.1)$ &
$2^{+1}_{-1}\times10^{-5}$ &
$<5$ &
$18.0^{+1.7}_{-1.1}$ &
$1.5^{+0.2}_{-0.2}\times10^{-1}$ &
$6.0^{+4.0}_{-2.0}\times10^{-3}$ &
$1.2^{+0.2}_{-0.2}$ &
$531^{+53}_{-68}$ &
$4.9^{+0.3}_{-0.4}$ &
$0.2$ &
${\scriptstyle(1.1^{+0.1}_{-0.1},0.9^{+0.1}_{-0.1},1.0^{+0.1}_{-0.1})}$ \\
$\mathrm{HD~}{\mathrm{Ring\ 2}}^{A}~(99.7)$ &
$2^{+3}_{-1}\times10^{-3}$ &
$1148^{+1607}_{-657}$ &
$15.5^{+1.4}_{-0.9}$ &
$4.7^{+1.0}_{-0.8}\times10^{-2}$ &
$9.0^{+11.0}_{-4.0}\times10^{-1}$ &
$0.6^{+0.1}_{-0.1}$ &
$108^{+21}_{-16}$ &
$4.4^{+0.3}_{-0.4}$ &
$2.6$ &
${\scriptstyle(0.9^{+0.1}_{-0.1},1.0^{+0.1}_{-0.1},1.0^{+0.1}_{-0.1})}$ \\
$\mathrm{HD~}{\mathrm{Ring\ 2}}^{B}~(99.7)$ &
$2^{+1}_{-1}\times10^{-5}$ &
$6^{+3}_{-3}$ &
$13.2^{+1.2}_{-1.5}$ &
$3.8^{+0.7}_{-0.5}\times10^{-1}$ &
$1.3^{+0.2}_{-0.5}\times10^{-2}$ &
$1.7^{+0.2}_{-0.2}$ &
$703^{+160}_{-100}$ &
$3.5^{+0.3}_{-0.2}$ &
$1.5$ &
${\scriptstyle(1.0^{+0.1}_{-0.1},0.9^{+0.1}_{-0.1},1.0^{+0.1}_{-0.1})}$ \\
\multicolumn{11}{c}{\rule{\linewidth}{0.4pt}} \\
$\mathrm{LkCa~}{\mathrm{Ring\ 1}}^{A}~(67.6)$ &
$3^{+1}_{-1}\times10^{-2}$ &
$2017^{+647}_{-503}$ &
$17.9^{+1.3}_{-1.1}$ &
$1.5^{+0.2}_{-0.2}\times10^{-2}$ &
$3.9^{+1.5}_{-1.1}\times10^{-1}$ &
$0.6^{+0.1}_{-0.1}$ &
$93^{+13}_{-11}$ &
$13.0^{+1.1}_{-1.0}$ &
$2.2$ &
${\scriptstyle(0.9^{+0.1}_{-0.1},0.9^{+0.1}_{-0.1},1.0^{+0.1}_{-0.1})}$ \\
$\mathrm{LkCa~}{\mathrm{Ring\ 1}}^{B}~(67.6)$ &
$2^{+2}_{-1}\times10^{-4}$ &
$<15$ &
$15.2^{+0.6}_{-0.9}$ &
$9.4^{+1.2}_{-0.7}\times10^{-2}$ &
$2.2^{+2.7}_{-1.0}\times10^{-3}$ &
$0.9^{+0.1}_{-0.1}$ &
$607^{+57}_{-40}$ &
$14.6^{+2.6}_{-1.6}$ &
$2.8$ &
${\scriptstyle(0.7^{+0.1}_{-0.1},0.8^{+0.1}_{-0.1},1.1^{+0.1}_{-0.1})}$ \\
$\mathrm{LkCa~}{\mathrm{Ring\ 2}}^{A}~(100.5)$ &
$2^{+1}_{-1}\times10^{-3}$ &
$310^{+153}_{-131}$ &
$12.2^{+0.7}_{-0.5}$ &
$3.6^{+2.2}_{-0.9}\times10^{-2}$ &
$1.1^{+0.4}_{-0.4}\times10^{-1}$ &
$0.8^{+0.4}_{-0.2}$ &
$189^{+85}_{-38}$ &
$9.0^{+1.1}_{-1.2}$ &
$1.0$ &
${\scriptstyle(1.1^{+0.1}_{-0.1},1.0^{+0.1}_{-0.1},1.0^{+0.1}_{-0.1})}$ \\
$\mathrm{LkCa~}{\mathrm{Ring\ 2}}^{B}~(100.5)$ &
$1^{+1}_{-1}\times10^{-4}$ &
$ <26 $ &
$10.0^{+0.9}_{-0.6}$ &
$1.6^{+0.2}_{-0.3}\times10^{-1}$ &
$5.0^{+4.0}_{-3.0}\times10^{-3}$ &
$0.8^{+0.1}_{-0.1}$ &
$975^{+121}_{-164}$ &
$11.3^{+1.0}_{-1.0}$ &
$1.3$ &
${\scriptstyle(1.1^{+0.1}_{-0.1},1.0^{+0.1}_{-0.1},1.0^{+0.1}_{-0.1})}$ \\
\enddata
\end{deluxetable*}

\begin{figure*}
\includegraphics[width=\textwidth]{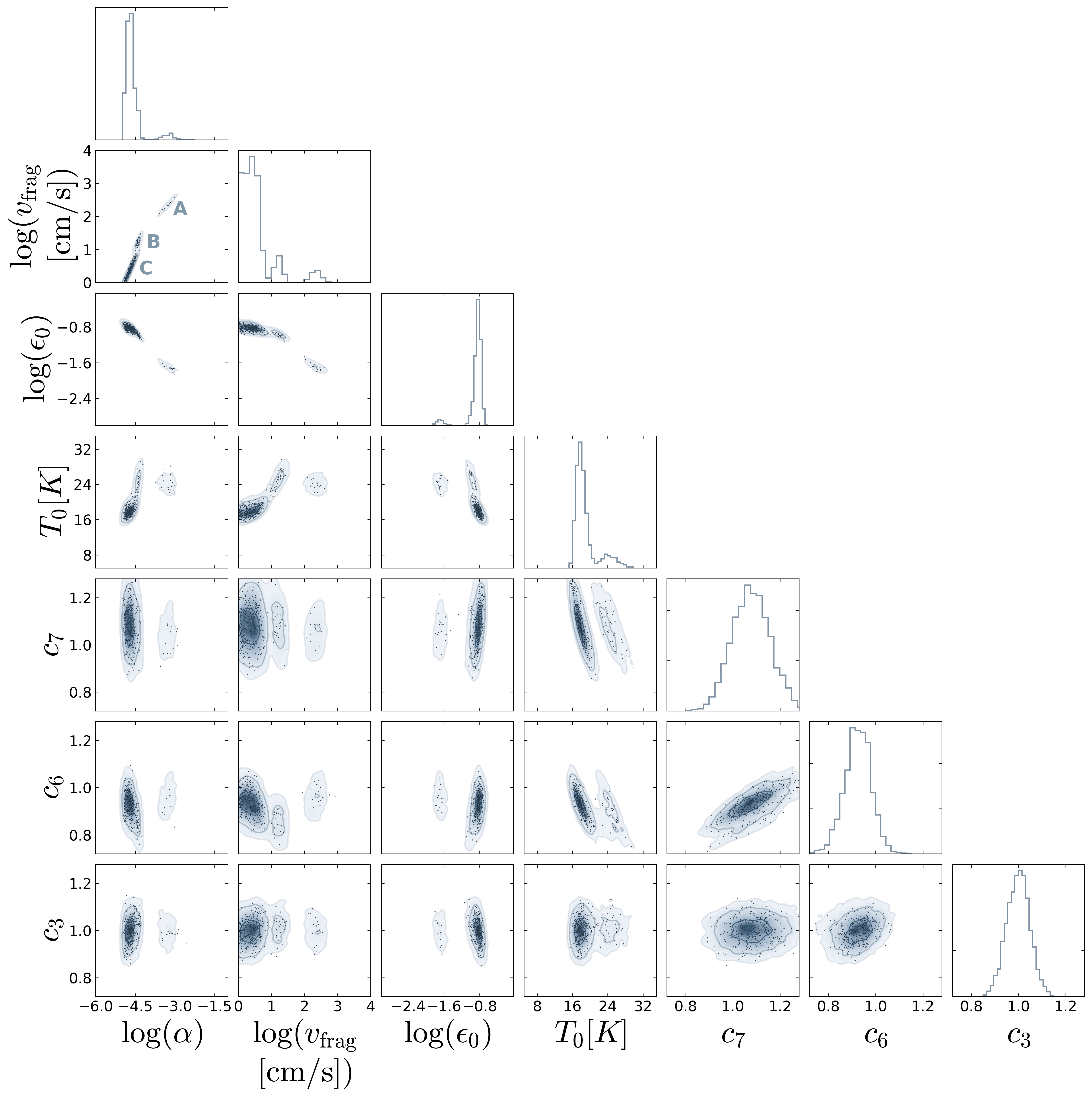}
\caption{
Posterior distributions for the ring 1 of HD 163296. Contours show the 68\% , 95\%, and 99\% credible regions of the inferred parameters
$(\alpha, v_{\mathrm{frag}}, \epsilon_0, T_0)$ and the multiplicative calibration factors $(c_7,c_6,c_3)$ (see definitions in Eq.~\ref{eq:likelihood}) for ALMA Bands~7, 6, and 3.
Three posterior clusters are identified due to model degeneracies, and the three branches are labeled A, B, and C in the $\alpha$--$v_{\rm frag}$ panel. Uniform priors are adopted in the plotted parameter space, and the axis limits indicate the lower and upper bounds of the adopted priors.
}
\label{fig:para_corner}
\end{figure*}

\subsection{Posterior Distributions and Parameter Degeneracies}
\label{subsec:posterior_degeneracies}

The corresponding posterior constraints and fitted ring properties are summarized in Table~\ref{tab:ALMA_fits}.
A representative example of the inferred posterior distributions for the fitted parameters of HD~163296 Ring~1 is shown in Figure~\ref{fig:para_corner}. The figure clearly reveals strong parameter correlations and the three distinct posterior clusters reported in Table~\ref{tab:ALMA_fits}. The inferred posterior distributions are generally degenerate and exhibit a multi-mode pattern. Specifically, $\alpha$ and $v_{\rm frag}$ are tightly correlated, whereas $\epsilon_0$ and $T_0$ follow different degeneracy directions depending on the mode.

This bimodality is primarily driven by the role of the maximum grain size in shaping the multi-wavelength spectral energy distribution (SED). One class of solutions favors very small grains ($a_{\max}\sim10^{-4}$--$10^{-3}\,\mathrm{cm}$), low turbulence, and a relatively high dust-to-gas ratio, whereas the other prefers much larger grains ($a_{\max}\sim0.1$--$1\,\mathrm{cm}$) together with higher $\alpha$ and $v_{\rm frag}$ in order to reproduce the observed spectral slopes. We therefore interpret the two clusters as physically distinct solutions permitted by the current data, rather than as a single uniquely determined parameter set.

\subsection{Gas-Dependent Constraints on $\alpha$ and $v_{\rm frag}$}
\label{subsec:alpha_vfrag_constraints}
The multi-wavelength fits provide statistical constraints on $\alpha$ and $v_{\rm frag}$ within our dust-ring model, but their absolute values remain coupled to the adopted gas prescription. In the fiducial fits, $\Sigma_{\rm g,0}$ and $w_{\rm p}$ are fixed to representative values, but the inferred solutions admit simple scaling relations: at fixed $\Sigma_{\rm g,0}$, $w_{\rm p}\propto \alpha^{-1/2}\propto v_{\rm frag}^{-1}$, whereas at fixed $w_{\rm p}$, $\Sigma_{\rm g,0}\propto \alpha^{-1}\propto v_{\rm frag}^{-1}$. The correlated posterior structure between $\alpha$ and $v_{\rm frag}$ reflects these dependencies, indicating that both parameters are sensitive to the assumed gas conditions rather than uniquely determined by the continuum data alone. We therefore regard the inferred $\alpha$ and $v_{\rm frag}$ as gas-dependent reference estimates, and discuss the robustness of the main conclusions to these scalings in Appendix~\ref{appendix:Robustness}.

Most of the low-$\alpha$, low-$v_{\rm frag}$ solutions should be interpreted with additional caution. These branches are boundary-limited: their posteriors reach the lower prior boundary adopted for the fragmentation velocity, $v_{\rm frag}=1~{\rm cm~s^{-1}}$. These solutions are therefore not spurious, although they can carry substantial posterior weight and similar degenerate low-$v_{\rm frag}$ branches also appear in our mock-recovery tests (Appendix~\ref{Appendix:Mock}; Table~\ref{tab:sim_fits}). They should be regarded as prior-truncated limiting branches rather than direct measurements of the physical fragmentation velocity.

Nevertheless, the turbulence levels favored by our fits are broadly compatible with the low-to-moderate mid-plane turbulence typically inferred for protoplanetary disks. Direct line-broadening measurements provide stringent upper limits in some systems (e.g., \citealt{Flaherty_2015,Flaherty_2017}, or see reviews by \citealt{Lesur2023PPVII,Rosotti2023}), meanwhile indirect constraints from vertical structure, thermochemical modeling, ring dust reservoirs, and planet-disk interaction simulations generally further support weak effective viscosity \citep[e.g.,][]{Dong_2017, Liu_2018,Zhang_2018,Doi_2021,Liuyao_2022,Garrido-Deutelmoser2023,Pizzati_2023,Lee_2024,Jiang_2024}.

The fragmentation thresholds implied by the different posterior modes should not be over-interpreted as direct material measurements. The higher-$v_{\rm frag}$ branches typically favor values of a few to $\sim10~\mathrm{m\,s^{-1}}$, and in some cases approach $\sim20~\mathrm{m\,s^{-1}}$. These values are above the commonly adopted fiducial value of $\sim 1\,\mathrm{m\,s^{-1}}$, but may still be physically plausible if the relevant grains are icy and porous: collision simulations have suggested that aggregate growth can survive impact velocities up to several tens of $\mathrm{m\,s^{-1}}$ under favorable conditions \citep{wada_2009}. The low-$v_{\rm frag}$ branches, on the other hand, are often prior-truncated limiting solutions rather than measurements of a specific fragmentation threshold. Interpreting these modes may require extending the dust-evolution framework to include additional collisional outcomes, such as bouncing\citep{Dominik_2023,Tong_2026}.

The inferred mid-plane temperatures remain in a plausible range of $\sim 10$--$25$ K, while the dust-to-gas ratio at the pressure maximum, $\epsilon_0$, differs more strongly between the two posterior modes. The large-grain solutions generally favor $\epsilon_0\sim10^{-2}$--$10^{-1}$, whereas the small-grain solutions require much higher local dust concentrations, in some cases approaching unity. This again highlights that the current data admit physically distinct solutions rather than a single uniquely determined parameter set.

\textcolor{red}{The small-grain solutions with high dust-to-gas ratios may also be susceptible to the streaming instability (SI). Although small Stokes numbers generally make strong SI clumping more difficult, the inferred large $\epsilon_0$ values, together with the very low turbulence levels in these solutions, can lead to substantial midplane dust enrichment and therefore conditions favorable for SI \citep{LiYoudin2021,Lim2024}. However, the applicability of standard SI thresholds to our models is uncertain because the dust is confined within a pressure bump, where the pressure gradient and radial drift velocity vary spatially. Dedicated pressure-bump simulations have shown that conventional local SI criteria are not sufficient to determine whether strong clumping occurs, particularly for small grains \citep{CarreraSimon2022,XuBai2022}. 
Observationally, there is some evidence that planetesimal formation could explain the observed optical depths if the SI maintains the midplane dust‑to‑gas ratio at unity \citep{Stammler2019}. This potential effect of the SI on ring morphology warrants more detailed investigation in the future.
}

\subsection{Derived Optical Depths, Dust Masses, and Intrinsic Widths}
\label{subsec:derived_quantities}

In addition to the directly sampled parameters, Table~\ref{tab:ALMA_fits} reports several derived ring properties constrained by the multi-wavelength fits. The peak absorption optical depths at 1.3~mm are generally of order unity, $\tau_0^{\rm 1.3mm}\simeq0.55$--$1.66$, indicating that the Band~6 emission from the ring centers is close to optically thick in all four fitted rings. Within this multi-wavelength fitting framework, the inferred optical depths and total dust masses remain relatively stable against changes in the assumed gas prescription, as shown by the gas-property tests in Appendix~\ref{appendix:Robustness}.

The inferred total dust masses are relatively large, ranging from tens to nearly $10^3\,M_{\oplus}$ across the posterior modes, where $M_{\oplus}$ is the Earth mass. The large-grain branches give dust masses broadly comparable to conventional estimates \citep{Manara_2023}, whereas the small-grain branches require substantially higher dust reservoirs. This behavior suggests that dust masses inferred from multi-wavelength ring modeling can be higher than simple estimates that neglect size-distribution, optical-depth, and resolution effects.

The inferred intrinsic dust-ring widths, represented by $w_{\epsilon}$, are typically a few to $\sim15$~au. These values are comparable to, or smaller than, the physical beam sizes of the fitted ALMA data, indicating that finite angular resolution remains important for interpreting the measured radial profiles. Apparent ring widths measured directly from individual wavelength profiles can also be affected by optical depth and scattering, and therefore should not be interpreted as direct measurements of the underlying dust distribution without forward modeling. Interpreting multi-wavelength ring widths thus requires a joint treatment of dust evolution, radiative transfer, and resolution.

We also tested the very recent disk continuum emission model of \citet{Kitade_Kataoka_2026} as an alternative to the Eddington, isotropic scattering, and  the two-stream approximations of \citet{zhu_2019}. Using this prescription shifts the solutions toward somewhat smaller $\alpha$ and $v_{\rm frag}$, and also lowers the inferred $\epsilon_0$, optical depth, and dust mass. These changes are typically at the normalization level, of order a factor of $\sim0.8$, rather than an order-of-magnitude difference, and they do not qualitatively alter the posterior structure or the main conclusions. We therefore retain the more mature and computationally efficient prescription \citep{zhu_2019} as the fiducial intensity model in this work.

\section{Conclusion}\label{sec:con}


Most interpretations of dust ring width in protoplanetary disks adopt a single-species description with a fixed representative grain size, and therefore neglect grain growth and the evolution of the size distribution. 
This simplification limits their ability to self-consistently connect ring observations across different ALMA bands.

We address this issue with a Bayesian inference framework that builds on the growth-fragmentation-trapping model of \citet{Yang_2025} and jointly incorporates radiative transfer, including scattering, and finite angular resolution. Applying this framework to rings in HD~163296 and LkCa~15, we show that the physical dust-ring model can simultaneously reproduce the observed multi-wavelength ALMA ring profiles by explicitly forward-modeling radiative-transfer and limited-resolution effects. 

Our results show that intrinsically narrow dust rings naturally account for the broader, wavelength-dependent widths recovered in the observations. More importantly, the framework provides a statistical path from direct observables to posterior constraints on key dust-evolution parameters, such as turbulence strength $\alpha$ and fragmentation velocity $v_{\rm frag}$, as well as derived quantities such as dust size $a_{\rm max}$, optical depth and dust mass  in a self-consistent way. 
\textcolor{red}{The fits reveal a common degeneracy across the rings: most rings admit both a low-$\alpha$, low-$v_{\rm frag}$ branch with small grains and high dust-to-gas ratios, and a higher-$\alpha$, higher-$v_{\rm frag}$ branch with larger grains and lower dust-to-gas ratios. Typical low-$\alpha$ branches have $\alpha\sim10^{-5}$--$10^{-4}$, $v_{\rm frag}$ of a few cm~s$^{-1}$ level, and $a_{\max}\sim10^{-3}$--$10^{-2}$ cm, whereas the higher-$\alpha$ branches reach $\alpha\sim10^{-3}$--$10^{-2}$, $v_{\rm frag}$ of a few -- $20~{\rm m~s^{-1}}$, and $a_{\max}\sim0.1$--$1$ cm.} 

This capability opens a more direct way to connect multi-wavelength continuum observations with the physical processes governing dust growth and trapping.
The accompanying open-source package \texttt{dring} provides a reusable implementation of this workflow for applying the same inference framework to other multi-wavelength dust-ring datasets.

Future longer-wavelength, high-resolution observations -- including ALMA Band~1, VLA Q/Ka Band and ultimately facilities, such as the ngVLA and Square Kilometre Array (SKA) -- will also be crucial for detecting (or ruling out) the predicted signatures and for tightening constraints on dust growth and trapping in protoplanetary disks \citep{Ricci2018,Li2019,Ilee2020,Li2020,Wu2024SKA,Wu-etal.2026-SKA,Garufi-etal2026-SKA}.

\section*{Acknowledgment}

\textcolor{red}{We thank the anonymous referee for a constructive report that helped improve the clarity of this manuscript.}
This research was supported by the funding from the National SKA Program of China under Grant No. 2025SKA0120100, the National Natural Science Foundation of China (grants 12373070, and 12192223). R.D. is supported by the Fundamental Research Funds for the Central Universities, Peking University. Y.W. acknowledges the EACOA Fellowship awarded by the East Asia Core Observatories Association. We also acknowledge the Chinese Center for Advanced Science and Technology for hosting the Proto-planetary Disk and Planet Formation Summer School in 2022,
organized by Xue-Ning Bai and Ruobing Dong,  which inspired the ideas underlying this work. The calculations have made use of the High Performance Computing Resource in the Core Facility for Advanced Research Computing at Shanghai Astronomical Observatory. 

Softwares: \texttt{Numpy} \citep{vanderWalt2011}, \texttt{Scipy} \citep{Virtanen2020}, \texttt{Matplotlib} \citep{Hunter2007}, \texttt{RADMC-3D} \citep{Radmc}, \texttt{CASA} \citep{CASA}, \texttt{UltraNest}\citep{ultranest1}.

\begin{appendix}

\section{Observational Fitting Setups}
\label{appendix:fitting_setups}
This appendix summarizes the observational setup and profile-extraction procedure used for the real ALMA rings. We describe the de-projection, azimuthal averaging, and effective angular resolution in Appendix~\ref{app:deprojection}, the adopted priors in Appendix~\ref{app_prior}, and the source-specific fitting choices for HD~163296 and LkCa~15 in Appendices~\ref{app:hd163296_setup} and~\ref{app:lkca15_setup}, respectively.

\subsection{Profile Extraction and Effective Angular Resolution}
\label{app:deprojection}
For each continuum image, we extract one-dimensional radial intensity profiles
using a common de-projection and azimuthal-averaging procedure. Sky-plane
offsets are first computed relative to the image center,
rotated into the disk major/minor-axis frame using the adopted position angle,
and de-projected by the disk inclination to define the cylindrical radius
\[
r = \sqrt{x_{\rm maj}^2 + (y_{\rm min}/\cos i)^2}.
\]
The images are then averaged in concentric annuli of de-projected radius.

The image rms, \(\sigma_{\rm rms,out}\), is estimated from an outer
emission-free annulus. The uncertainty on
the mean intensity in each radial bin is scaled by the number of independent
beam elements in that annulus,
\[
\sigma_{\rm rms}(r) =
\frac{\sigma_{\rm rms,out}}{\sqrt{N_{\rm beam}(r)}} ,
\]
where \(N_{\rm beam}(r)=N_{\rm pix}(r)/A_{\rm beam,pix}\), \(N_{\rm pix}(r)\)
is the number of image pixels falling in the radial annulus, and
\(A_{\rm beam,pix}\) is the synthesized beam area expressed in pixel units.
The same procedure is applied to all simulated and observed datasets used in this work.

For each image, we characterize the angular resolution by the effective beam size
\begin{equation}
\theta_{\rm res}= \sqrt{{\rm FWHM}_{\rm maj}\,{\rm FWHM}_{\rm min}},
\end{equation}
where ${\rm FWHM}_{\rm maj}$ and ${\rm FWHM}_{\rm min}$ are the major and minor axes of the restoring beam.
This $\theta_{\rm res}$ corresponds to the effective angular resolution used in Section~\ref{subsec:RT_Res}, i.e., the FWHM of the equivalent Gaussian beam, and sets the Gaussian convolution kernel used to account for finite angular resolution.

\subsection{Priors and Calibration Factors}
\label{app_prior}
For all fits in this work, including the mock-recovery tests, we adopt the same baseline priors for the physical parameters,
\begin{equation}
\begin{aligned}
P(\log_{10}\alpha) &= \mathcal{U}(-6,\,-1),\\
P\!\left[\log_{10}\left(\frac{v_{\rm frag}}{{\rm cm~s^{-1}}}\right)\right]
&= \mathcal{U}(0,\,4),\\
P(\log_{10}\epsilon_0) &= \mathcal{U}(-3,\,0),\\
P(T_0 /~{\rm K}) &= \mathcal{U}(5,\,25).
\end{aligned}
\end{equation}
Here $\mathcal{U}(a,b)$ denotes a uniform, or flat, prior between $a$ and $b$.
The only exception is HD~163296 Ring~1, for which the upper bound of the temperature prior is extended to $T_0=30~{\rm K}$.

For the absolute flux calibration, following the ALMA Cycle~12 Technical Handbook, we adopt fractional uncertainties
$f_{\rm cal}=0.1$, $0.1$, and $0.05$ for ALMA Bands~7, 6, and 3, respectively. For each wavelength, we assign a flat prior
\begin{equation}
P(c_{\lambda_n}) =
\mathcal{U}\left(1-3f_{{\rm cal},\lambda_n},\,1+3f_{{\rm cal},\lambda_n}\right),
\end{equation}
The posterior is sampled using \texttt{UltraNest} \citep{ultranest1}. The resulting posterior distributions are used to constrain the free parameters as well as the corresponding derived structural quantities.

\subsection{HD~163296}
\label{app:hd163296_setup}

We fit the two prominent rings in HD~163296 using azimuthally averaged radial intensity profiles extracted from the CLEAN images presented by \citet{Guidi_2022}. The profile extraction follows the de-projection and azimuthal-averaging procedure described in Appendix~\ref{app:deprojection}.

Following the resolution treatment in Section~\ref{subsec:RT_Res}, the fitted bands, effective angular resolutions, and adopted stellar/disk parameters are
\begin{equation}
\begin{aligned}
(\lambda_{7},\lambda_{6},\lambda_{3})
&=(0.89,\,1.30,\,3.20)~{\rm mm},\\
(\theta_{{\rm res},7},\theta_{{\rm res},6},\theta_{{\rm res},3})
&=(0.0667,\,0.0428,\,0.0783)~{\rm arcsec},\\
(d,\,i,\,M_\star)
&=(101.2~{\rm pc},\,46^\circ,\,1.9\,M_\odot).
\end{aligned}
\end{equation}
Here $\theta_{\rm res}$ is the effective angular resolution computed from the ALMA restoring beam as described in Appendix~\ref{app:deprojection}.

For the two rings, the fixed ring centers, sampled radii used in the likelihood evaluation, and adopted gas-profile parameters are shown below. \textcolor{red}{We also list the grain size corresponding to ${\rm St}_{\max}=1$, computed using the default DSHARP dust composition (i.e., the bulk density of the dust particles $\rho_{\rm s}$), to indicate the upper boundary imposed by the ${\rm St}_{\max}<1$ condition.}
\begin{equation}
\begin{aligned}
{\rm Ring~1:}\quad
&r_0=67.0~{\rm au},\quad
\{r_m\}=[64,\,67,\,71]~{\rm au},\\
&w_{\rm p}=14.4~{\rm au},\quad
\Sigma_{\rm g,0}=21~{\rm g~cm^{-2}},\\
&\textcolor{red}{a({\rm St}_{\max}=1)=8.0~{\rm cm}},\\[0.5ex]
{\rm Ring~2:}\quad
&r_0=99.5~{\rm au},\quad
\{r_m\}=[95,\,99.5,\,104]~{\rm au},\\
&w_{\rm p}=23.2~{\rm au},\quad
\Sigma_{\rm g,0}=10~{\rm g~cm^{-2}},\\
&\textcolor{red}{a({\rm St}_{\max}=1)=3.8~{\rm cm}}.
\end{aligned}
\end{equation}
The ring centers are obtained from our own initial fit to the observed profiles. The bump widths are adopted from the results reported by \citet{rosotti_2020}, while the gas surface densities are set based on previous gas-structure estimates \citep{Booth_2019,Doi_2021}. The same uncertainty prescription as in Appendix~\ref{Appendix:Mock} is adopted, combining the profile uncertainty $\sigma_I$ with the band-dependent calibration term. The physical parameters sampled in the Bayesian inference are $\alpha$, $v_{\rm frag}$, $T_0$, and $\epsilon_0$.

\subsection{LkCa~15}
\label{app:lkca15_setup}

We fit the two prominent rings in LkCa~15 using azimuthally averaged radial intensity profiles extracted from the CLEAN images presented by \citet{Sierra_2025}. As for HD~163296, the profiles are measured with the same de-projection and azimuthal-averaging procedure adopted for the mock datasets.

Following the resolution treatment in Section~\ref{subsec:RT_Res}, the fitted bands, effective angular resolutions, and adopted stellar/disk parameters are
\begin{equation}
\begin{aligned}
(\lambda_{7},\lambda_{6},\lambda_{3})
&=(0.882,\,1.34,\,3.08)~{\rm mm},\\
(\theta_{{\rm res},7},\theta_{{\rm res},6},\theta_{{\rm res},3})
&=(0.06,\,0.06,\,0.06)~{\rm arcsec},\\
(d,\,i,\,M_\star)
&=(157.2~{\rm pc},\,50.1^\circ,\,1.2\,M_\odot).
\end{aligned}
\end{equation}
Here $\theta_{\rm res}$ has the same meaning as in Appendix~\ref{app:deprojection}: the effective angular resolution computed from the ALMA restoring beam.

For the two rings, the fixed ring centers, sampled radii used in the likelihood evaluation, and adopted gas-profile parameters are shown below. \textcolor{red}{We also list the grain size corresponding to ${\rm St}_{\max}=1$, computed using the default DSHARP dust composition, to indicate the upper boundary imposed by the ${\rm St}_{\max}<1$ condition.}
\begin{equation}
\begin{aligned}
{\rm Ring~1:}\quad
&r_0=67.6~{\rm au},\quad
\{r_m\}=[61,\,68,\,75]~{\rm au},\\
&w_{\rm p}=8.0~{\rm au},\quad
\Sigma_{\rm g,0}=23~{\rm g~cm^{-2}},\\
&\textcolor{red}{a({\rm St}_{\max}=1)=8.7~{\rm cm}},\\[0.5ex]
{\rm Ring~2:}\quad
&r_0=100.5~{\rm au},\quad
\{r_m\}=[96.5,\,102.5,\,108.5]~{\rm au},\\
&w_{\rm p}=15.0~{\rm au},\quad
\Sigma_{\rm g,0}=12~{\rm g~cm^{-2}},\\
&\textcolor{red}{a({\rm St}_{\max}=1)=4.6~{\rm cm}}.
\end{aligned}
\end{equation}
The ring centers are obtained from our own initial fit to the observed profiles. The bump widths are adopted following the method of \citet{rosotti_2020} with $^{12}\rm CO$ rotation-curve constraints \citep{Stadler_2025}, and the gas surface densities are fixed based on gas-chemical constraints \citep{Sierra_2025,Jin_2019}. The physical parameters sampled in the Bayesian inference are also $\alpha$, $v_{\rm frag}$, $T_0$, and $\epsilon_0$. 

\section{Mock Observations and Recovery Tests}
\label{Appendix:Mock}
\begin{figure*}
\includegraphics[width=\textwidth]{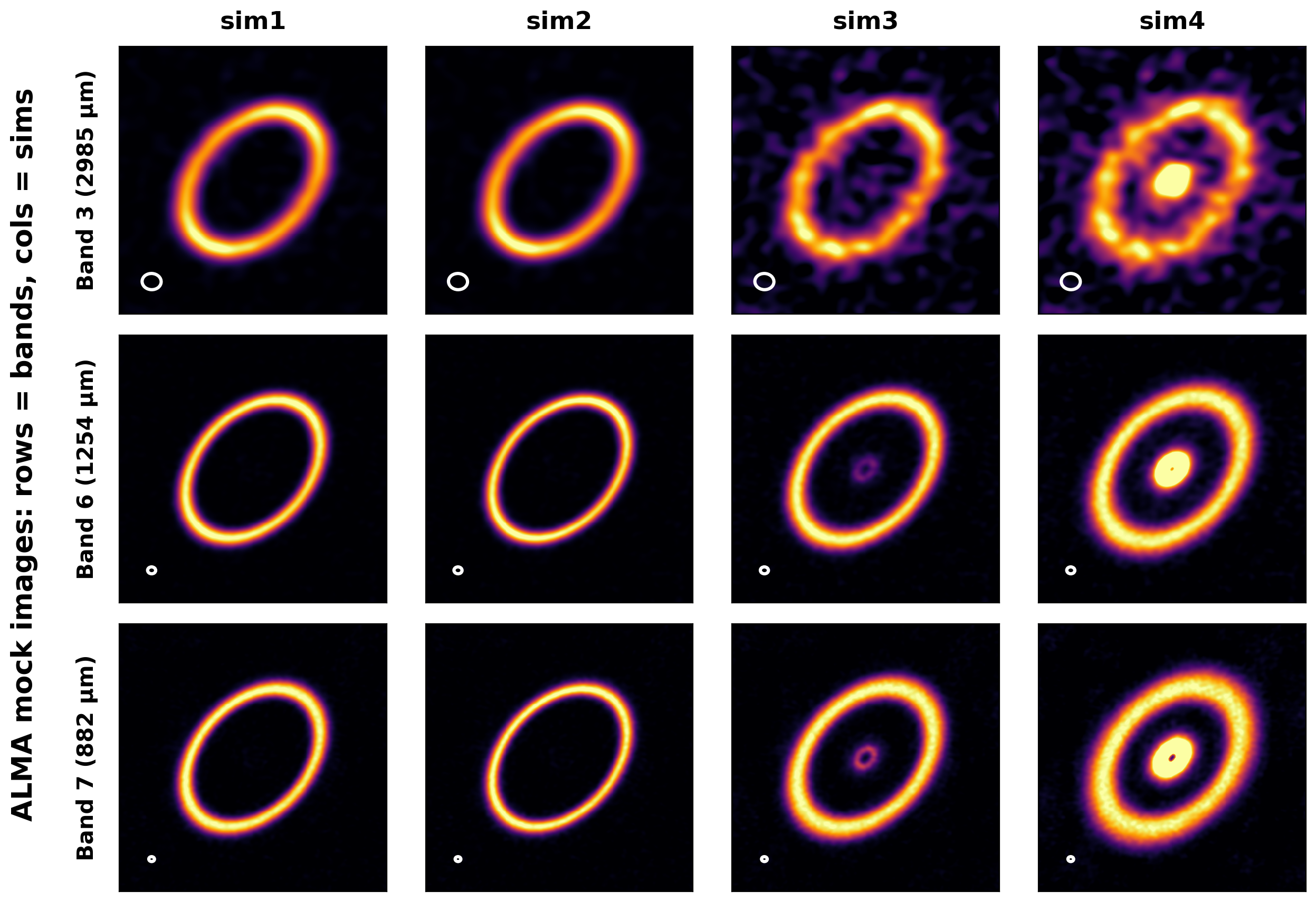}
\caption{
CASA imaging results for the mock ALMA continuum observations used for pipeline validation.
Rows show the four simulated dust-ring models (Sim~1--Sim~4), and columns correspond to ALMA Bands~3, 6, and~7 (2.985, 1.254, and 0.882 mm, respectively).
For each simulation, mock visibilities are generated from the corresponding RADMC--3D sky model, with thermal noise included, using a compact plus extended array setup to reduce spatial filtering while retaining high angular resolution.
All images are reconstructed with the same CLEAN settings (Briggs weighting with robust $=0.5$).
The restoring beam for each image is shown in the lower-left corner.
This figure serves as a qualitative illustration of wavelength-dependent radiative-transfer and resolution effects and is not used directly in the Bayesian inference.
}
\label{fig:casa_image}
\end{figure*}
We construct four synthetic dust-ring models to validate the inference framework under realistic multi-wavelength ALMA observing conditions. All \texttt{dustpy} simulations follow the fiducial dust-trapping setup of \citet{Yang_2025}, with a Gaussian gas-pressure bump centered at $r_0=100\,\mathrm{au}$ and dust evolution governed by radial drift, turbulent diffusion, coagulation, and fragmentation. Relative to the fiducial model, we adopt a lower gas surface density normalization, $\Sigma_{\rm g,0}=5~\mathrm{g~cm^{-2}}$, so that the mock datasets span both optically thin and moderately thick regimes. The four simulation cases are generated by varying $\alpha$ and $v_{\rm frag}$ while keeping all other parameters fixed; the adopted combinations are listed in the rows labeled ``Setup'' in Table~\ref{tab:sim_fits}.

This appendix first shows the mock ALMA observation setup in Section~\ref{app:casa_mock} and summarize the corresponding fitting results in Section~\ref{app:mock_fitting}.



\subsection{Mock ALMA Observations}
\label{app:casa_mock}

We use the RADMC--3D radiative-transfer output \citep{Radmc} as the intrinsic sky brightness model 
and input it into CASA to generate mock ALMA continuum visibilities \citep[][version 6.7.0-31]{CASA}.
To construct the three-dimensional density field required for radiative transfer, each grain-size bin is assigned a Gaussian vertical distribution\citep{Stammler_2022},
\begin{equation}
\begin{aligned}
\rho_{\rm d}(r,z,a)
&=
\frac{\Sigma_{\rm d}(r,a)}{\sqrt{2\pi}H_{\rm d}(r,a)}
\exp\!\left[-\frac{z^2}{2H_{\rm d}^2(r,a)}\right],\\
H_{\rm d}(r,a)
&=
H_{\rm g}(r)
\left(1+\frac{\mathrm{St}(r,a)}{\alpha}\right)^{-1/2}.
\end{aligned}
\end{equation}
The mock disk is placed at the sky position of HD~163296, with $(\alpha,\delta)_{\rm J2000}=(269.0887008^\circ,-21.9560755^\circ)$, and adopts a distance of 101.2~pc, an inclination of $46^\circ$, and a position angle of $133^\circ$. We simulate continuum observations in ALMA Bands~3, 6, and~7, with central wavelengths of 2.985, 1.254, and 0.882 mm, respectively, using a total bandwidth of 7.5~GHz in multi-frequency synthesis mode.

To recover both the narrow ring width and the larger-scale annular emission, we use a compact plus extended array setup. Specifically, the mock visibilities are generated with the CASA configurations \texttt{alma.cycle11.4.cfg} and \texttt{alma.cycle11.7.cfg}, with on-source times of 0.46 and 2 hr, respectively. 
This combination is adopted to reduce spatial filtering of the smooth ring component while retaining high angular resolution for the narrow ring structure. Thermal noise is included using the \texttt{tsys-atm} prescription, and all other observational settings are held fixed across wavelengths. 

The simulated visibilities are imaged with the CASA task \texttt{tclean} using multi-frequency synthesis and Briggs weighting with a robust parameter of 0.5. All imaging settings are kept fixed across wavelengths. Figure~\ref{fig:casa_image} shows the corresponding mock continuum images.

\subsection{Mock-Recovery Results}
\label{app:mock_fitting}

The mock datasets are fitted with the same Bayesian inference framework, uncertainty treatment, and priors used for the observational analysis, as summarized in Appendix~\ref{app_prior}. The likelihood is evaluated over a sparse set of radial points,
\begin{equation}
\begin{aligned}
{\rm Sim~1,\ Sim~2}: \quad
{r_m} &= [95,\,100,\,105]~{\rm au},\\
{\rm Sim~3,\ Sim~4}: \quad
{r_m} &= [92,\,100,\,108]~{\rm au}.
\end{aligned}
\end{equation}
They mirror the sparse likelihood sampling adopted for the observational data and use the same profile-extraction and width-measurement procedure as in the main text.

For the three observing bands considered here, following section \ref{app:deprojection}, the resulting effective angular resolutions are
\begin{equation}
(\theta_{{\rm res},7},\,\theta_{{\rm res},6},\,\theta_{{\rm res},3})
=(0.0658,\,0.0928,\,0.2188)~{\rm arcsec},
\end{equation}
listed in order of increasing wavelength. These values are used consistently in the resolution-aware modeling of the radial intensity profiles. For Sim~1 and Sim~2, the rings are only marginally resolved in Bands~7 and~6, and are unresolved in Band~3.

The resulting posterior constraints are summarized in Table~\ref{tab:sim_fits}. For each simulated dataset, at least one posterior mode remains reasonably consistent with the input setup, while multi-modal cases are correctly identified as such. However, the nested-sampling mode weights should not be interpreted as a reliable criterion for selecting the physically correct branch, since the input solution can appear in a lower-weight mode. Although modest offsets are present in some individual parameters, the relative trends among different combinations of $\alpha$ and $v_{\rm frag}$ are preserved. In particular, simulation setups with higher or lower turbulence and fragmentation thresholds are recovered in the correct qualitative order. 

The largest discrepancies occur for Sim~3 and Sim~4, especially in the inferred optical depth and temperature. These two mock datasets have lower signal-to-noise ratios, weaker constraints on the continuum ring width, and no strongly optically thick band, making $T_0$ and $\tau_0$ less well constrained. The other parameters and the qualitative ordering of the input models are nevertheless recovered reasonably well. 

These tests show that the framework captures the main ring properties and the dominant parameter trends relevant to this work, while naturally exposing the remaining degeneracies under realistic observational conditions. This remains possible even when the rings are not fully resolved. The framework jointly accounts for data taken at different angular resolutions without requiring the images to be artificially convolved to a common beam, thereby avoiding the loss of information from the higher-resolution data.

\begin{deluxetable*}{lcccccccccc}
\tablecaption{
Multi-mode Bayesian fits to simulated dust rings used for pipeline validation. The listed physical quantities and derived ring properties are defined in the same way as in Table~\ref{tab:ALMA_fits}.
For each mock dataset (Sim~1--Sim~4), the input parameters adopted in the simulations are listed under the rows ``Setup'', 
followed by the posterior constraints inferred from the resolution-aware Bayesian inference pipeline.
Quoted values correspond to posterior medians with 68\% credible intervals.
When the posterior distribution is multi-mode, distinct solutions (mode~A and mode~B) are reported separately, together with their nested-sampling mode weights $w_{\mathrm{mode}}$ indicating the relative posterior probability of each solution.
The recovered parameter values are generally consistent with the input parameters within uncertainties, demonstrating that the pipeline can robustly recover the underlying dust-evolution parameters from resolution-limited mock observations while correctly identifying physically distinct degenerate solutions.
}
\label{tab:sim_fits}
\tablehead{
\colhead{Source} &
\colhead{Mode($w_{\rm mode}$)} &
\colhead{$\alpha$} &
\colhead{$v_{\rm frag}$} &
\colhead{$T_0$} &
\colhead{$\epsilon_0$} &
\colhead{$a_{\max}$} &
\colhead{$\tau_{0}^{\rm 1.3mm}$} &
\colhead{$M_{\rm dust}$} &
\colhead{$w_{\epsilon}$} &
\colhead{$\chi^2_{\rm avg}$} \\
\colhead{$-$} &
\colhead{$-$} &
\colhead{$-$} &
\colhead{[cm s$^{-1}$]} &
\colhead{[K]} &
\colhead{$-$} &
\colhead{[cm]} &
\colhead{$-$} &
\colhead{[$M_{\oplus}$]} &
\colhead{[au]} &
\colhead{$-$} 
}
\colnumbers
\startdata
Sim~1 & Setup &
$1\times10^{-3}$&
$500$ &
$12.3$ &
$7.4\times10^{-2}$ &
$1\times10^{-1}$ & 
$0.6$&
$110$ &
$4.8$ &
$-$\\
~ & A($0.07$) &
$1^{+1}_{-1}\times10^{-3}$ &
$449^{+81}_{-57}$ &
$11.6^{+0.5}_{-0.5}$ &
$7.9^{+0.8}_{-0.7}\times10^{-2}$ &
$1.8^{+0.1}_{-0.3}\times10^{-1}$ &
$0.7^{+0.1}_{-0.1}$ &
$122^{+8}_{-6}$ &
$5.8^{+0.3}_{-0.3}$ &
$7.0$ \\
~ & B($0.93$) &
$2^{+1}_{-1}\times10^{-5}$ &
$<4$ &
$9.2^{+0.3}_{-0.3}$ &
$4.9^{+0.4}_{-0.3}\times10^{-1}$ &
$2.2^{+0.8}_{-0.6}\times10^{-3}$ &
$1.0^{+0.1}_{-0.1}$ &
$872^{+46}_{-44}$ &
$6.8^{+0.2}_{-0.2}$ &
$5.9$ \\
\multicolumn{11}{c}{\rule{\linewidth}{0.4pt}} \\  
Sim~2 & Setup &
$1\times10^{-3}$&
$800$ &
$12.3$ &
$1.0\times10^{-1}$ &
$4\times10^{-1}$ & 
$0.8$&
$111$  &
$3.0$ &
$-$\\
~ & A($0.54$) &
$2^{+1}_{-1}\times10^{-3}$ &
$1037^{+553}_{-349}$ &
$10.7^{+0.4}_{-0.4}$ &
$1.1^{+0.2}_{-0.1}\times10^{-1}$ &
$4.6^{+2.9}_{-1.3}\times10^{-1}$ &
$0.8^{+0.1}_{-0.1}$ &
$130^{+17}_{-13}$ &
$4.6^{+0.2}_{-0.2}$ &
$7.6$ \\
~ & B($0.46$) &
$1^{+1}_{-1}\times10^{-5}$ &
$<2$ &
$9.6^{+0.4}_{-0.3}$ &
$5.6^{+0.5}_{-0.5}\times10^{-1}$ &
$2.5^{+0.4}_{-0.4}\times10^{-3}$ &
$1.1^{+0.1}_{-0.1}$ &
$753^{+53}_{-43}$ &
$5.0^{+0.1}_{-0.2}$ &
$8.4$ \\
\multicolumn{11}{c}{\rule{\linewidth}{0.4pt}} \\  
Sim~3 & Setup &
$1\times10^{-3}$&
$200$ &
$11.6$ &
$3.4\times10^{-2}$ &
$2\times10^{-2}$ & 
$0.2$&
$97$ &
$12.0$&
$-$\\
~ & A($1$) &
$2^{+1}_{-1}\times10^{-3}$ &
$358^{+109}_{-51}$ &
$20.1^{+1.2}_{-1.2}$ &
$1.6^{+0.2}_{-0.3}\times10^{-2}$ &
$3.4^{+1.3}_{-0.5}\times10^{-2}$ &
$0.1^{+0.1}_{-0.1}$ &
$60^{+8}_{-12}$ &
$16.4^{+0.5}_{-0.5}$ &
$17.5$ \\
\multicolumn{11}{c}{\rule{\linewidth}{0.4pt}} \\  
Sim~4 & Setup &
$5\times10^{-3}$&
$500$ &
$12.3$ &
$1.8\times10^{-2}$ &
$3\times10^{-2}$ & 
$0.1$&
$64$ &
$21.6$&
$-$\\
~ &  A($1$) &
$7^{+4}_{-2}\times10^{-3}$ &
$621^{+375}_{-131}$ &
$18.0^{+1.6}_{-1.4}$ &
$1.3^{+0.3}_{-0.4}\times10^{-2}$ &
$3.4^{+3.1}_{-0.5}\times10^{-2}$ &
$0.1^{+0.1}_{-0.1}$ &
$68^{+15}_{-19}$ &
$27.2^{+2.7}_{-2.1}$ &
$11.8$ \\
\enddata
\end{deluxetable*}

\section{Robustness to Assumed Gas Properties and Grain-Size Slope}
\label{appendix:Robustness}

In this section we quantify how the inferred parameters respond to assumptions about the gas pressure bump and the grain-size distribution. Because the continuum profiles alone do not uniquely constrain $\Sigma_{\rm g,0}$ or $w_{\rm p}$, the absolute values of $\alpha$ and $v_{\rm frag}$ should be interpreted together with these assumed gas properties. We therefore vary $\Sigma_{\rm g,0}$, $w_{\rm p}$, \textcolor{red}{and the grain-size distribution slope $q$}, identify the dominant scaling degeneracies, and test whether the main conclusions about intrinsic ring widths and processed width--wavelength behavior remain stable under these variations.

Table~\ref{tab:robustness} presents these tests for HD~163296 Ring~1, which we use as a representative example. When the assumed gas surface density normalization $\Sigma_{\rm g,0}$ or pressure-bump width $w_{\rm p}$ is varied and the data are re-fitted, the posterior shifts mainly along correlated directions in parameter space while preserving a similar fit quality. Across these gas-property tests, the inferred maximum grain size $a_{\max}$ and the characteristic dust-to-gas ratio width $w_{\epsilon}$ remain approximately stable. This behavior is captured by the approximate scaling relations
\begin{align}
\left\{
\begin{aligned}
a_{\max} &\propto \Sigma_{\rm g}\,\mathrm{St}_{\rm frag}\propto\Sigma_{\rm g}\frac{v^2_{\rm frag}}{\alpha}, \label{eq:scalings_eqs}\\
w_{\epsilon} &\propto w_{\rm p} \sqrt{\frac{\alpha}{\mathrm{St}_{\rm frag}}} \propto w_{\rm p}\frac{\alpha}{v_{\rm frag}}.
\end{aligned}
\right.
\end{align}
The approximate stability of these two quantities is expected from the observables being fitted. The multi-wavelength SED requires a particular range of $a_{\max}$ to reproduce the spectral slopes, while the multi-wavelength continuum ring widths are primarily sensitive to the intrinsic ring width, which is set by $w_{\epsilon}$ in the dust-ring model. The observations therefore effectively constrain $a_{\max}$ and $w_{\epsilon}$ even when the assumed gas properties are varied.
Under the additional requirement that $a_{\max}$ and $w_{\epsilon}$ remain approximately fixed, 
these relations imply
\begin{align}
\left\{
\begin{aligned}
w_{\rm p} &\propto \alpha^{-1/2} \propto v_{\rm frag}^{-1}, \rm with\ \Sigma_{\rm g,0}\ fixed, \label{eq:gas_scalings}\\
\Sigma_{\rm g,0} &\propto \alpha^{-1} \propto v_{\rm frag}^{-1}, {\rm with}\ w_{\rm p}\ {\rm fixed}.
\end{aligned}
\right.
\end{align}
These relations provide a useful guide to the dominant degeneracy directions, but they should not be interpreted as exact or universal. In practice, changes in the assumed gas structure can also be partly absorbed by other fitted quantities, especially $T_0$ and, in some cases, $\epsilon_0$. The final inference must therefore rely on explicit re-fitting rather than on scaling arguments alone.

\textcolor{red}{We also test the sensitivity to the assumed grain-size distribution slope by repeating the fit with $q=2.5$, $3.0$, and $3.9$. As $q$ increases, the preferred solutions show a mild decrease in $\alpha$, a mild increase in $v_{\rm frag}$, and a slightly narrower intrinsic dust-ring width $w_{\epsilon}$ as shown in Table~\ref{tab:robustness}. The steeper distribution also places more dust mass in small grains and shifts the preferred solution toward larger $\epsilon_0$, $a_{\max}$, and total dust mass. Although the inferred dust-evolution parameters and dust mass can shift with the assumed $q$, the characteristic ring width remains within a narrow range of $\sim4$--$5$ au, and the main conclusion that the observed profiles can be reproduced by intrinsically narrow dust rings is unchanged.}

\begin{deluxetable*}{lcccccccccc}
\tablecaption{
Robustness tests for the inferred dust-ring parameters of HD~163296 ring~1 mode A. The listed physical quantities and derived ring properties are defined in the same way as in Table~\ref{tab:ALMA_fits}. For each case, we report the posterior medians and 68\% credible intervals, together with the nested-sampling mode weight $w_{\rm mode}$. Starting from the fiducial fit ("Fiducial"), we repeat the Bayesian inference while varying the assumed gas surface density normalization $\Sigma_{\rm g,0}$, the pressure-bump width $w_{\rm p}$, and \textcolor{red}{the grain-size distribution slope $q$}. 
\label{tab:robustness}}
\tablehead{
\colhead{Variable} &
\colhead{Mode($w_{\rm mode}$)} &
\colhead{$\alpha$} &
\colhead{$v_{\rm frag}$[cm/s]} &
\colhead{$T_0$ [K]} &
\colhead{$\epsilon_0$} &
\colhead{$a_{\max}$ [cm]} &
\colhead{$\tau_{0}^{\rm 1.3mm}$} &
\colhead{$M_{\rm dust}[M_{\oplus}]$} &
\colhead{$w_{\epsilon}$[au]} &
\colhead{$\mathrm{\chi}^2_{\rm avg}$}
}
\colnumbers
\startdata
Fiducial & A($0.05$) &
$5^{+2}_{-2}\times10^{-4}$ &
$214^{+83}_{-58}$ &
$24.2^{+1.6}_{-1.5}$ &
$2.1^{+0.4}_{-0.3}\times10^{-2}$ &
$1.7^{+0.7}_{-0.3}\times10^{-1}$ &
$0.8^{+0.1}_{-0.1}$ &
$68^{+9}_{-7}$ &
$4.4^{+0.4}_{-0.3}$ &
$0.9$ \\
\multicolumn{11}{c}{\rule{\linewidth}{0.4pt}} \\  
$\Sigma_{\rm g,0}\times5$ & 
A($0.36$) &
$1^{+1}_{-1}\times10^{-4}$ &
$45^{+18}_{-13}$ &
$24.2^{+1.7}_{-1.4}$ &
$4.1^{+0.8}_{-0.5}\times10^{-3}$ &
$1.7^{+0.7}_{-0.3}\times10^{-1}$ &
$0.8^{+0.2}_{-0.1}$ &
$68^{+8}_{-6}$ &
$4.5^{+0.4}_{-0.4}$ &
$0.9$ \\
$\Sigma_{\rm g,0}\times0.5$ & A($0.04$) &
$1^{+1}_{-1}\times10^{-3}$ &
$454^{+131}_{-131}$ &
$23.9^{+1.6}_{-1.3}$ &
$4.1^{+0.7}_{-0.5}\times10^{-2}$ &
$1.7^{+0.7}_{-0.3}\times10^{-1}$ &
$0.8^{+0.2}_{-0.1}$ &
$68^{+8}_{-4}$ &
$4.6^{+0.3}_{-0.4}$ &
$0.9$ \\
\multicolumn{11}{c}{\rule{\linewidth}{0.4pt}} \\  
$w_{\rm p}\times2$ & A($0.14$) &
$1^{+1}_{-1}\times10^{-4}$ &
$112^{+45}_{-28}$ &
$24.0^{+1.6}_{-1.3}$ &
$2.0^{+0.4}_{-0.3}\times10^{-2}$ &
$1.8^{+1.1}_{-0.3}\times10^{-1}$ &
$0.8^{+0.2}_{-0.1}$ &
$68^{+8}_{-5}$ &
$4.4^{+0.4}_{-0.4}$ &
$0.9$ \\
$w_{\rm p}\times0.5$ & A($0.03$) &
$3^{+1}_{-1}\times10^{-3}$ &
$533^{+185}_{-146}$ &
$24.1^{+1.5}_{-1.3}$ &
$1.9^{+0.4}_{-0.2}\times10^{-2}$ &
$1.7^{+0.7}_{-0.3}\times10^{-1}$ &
$0.7^{+0.1}_{-0.1}$ &
$67^{+6}_{-6}$ &
$5.4^{+0.6}_{-0.6}$ &
$1.0$ \\
\multicolumn{11}{c}{\rule{\linewidth}{0.4pt}} \\
\textcolor{red}{$q=2.5$} & \textcolor{red}{A($0.003$)} &
\textcolor{red}{$7.3^{+0.6}_{-1.5}\times10^{-4}$} &
\textcolor{red}{$196^{+27}_{-32}$} &
\textcolor{red}{$24.8^{+0.5}_{-1.9}$} &
\textcolor{red}{$1.5^{+0.3}_{-0.1}\times10^{-2}$} &
\textcolor{red}{$1.1^{+0.2}_{-0.0}\times10^{-1}$} &
\textcolor{red}{$0.8^{+0.2}_{-0.0}$} &
\textcolor{red}{$52^{+5}_{-4}$} &
\textcolor{red}{$5.0^{+0.3}_{-0.4}$} &
\textcolor{red}{$1.6$} \\
\textcolor{red}{$q=3.0$} & \textcolor{red}{A($0.01$)} &
\textcolor{red}{$5.9^{+2.6}_{-1.7}\times10^{-4}$} &
\textcolor{red}{$203^{+52}_{-48}$} &
\textcolor{red}{$24.6^{+1.6}_{-2.2}$} &
\textcolor{red}{$1.8^{+0.3}_{-0.3}\times10^{-2}$} &
\textcolor{red}{$1.3^{+0.2}_{-0.2}\times10^{-1}$} &
\textcolor{red}{$0.9^{+0.2}_{-0.2}$} &
\textcolor{red}{$57^{+6}_{-5}$} &
\textcolor{red}{$4.6^{+0.5}_{-0.4}$} &
\textcolor{red}{$1.4$} \\
\textcolor{red}{$q=3.9$} & \textcolor{red}{A($0.33$)} &
\textcolor{red}{$3.0^{+9.0}_{-2.0}\times10^{-4}$} &
\textcolor{red}{$247^{+685}_{-127}$} &
\textcolor{red}{$23.6^{+1.6}_{-1.2}$} &
\textcolor{red}{$3.3^{+0.5}_{-0.4}\times10^{-2}$} &
\textcolor{red}{$3.9^{+10.6}_{-1.9}\times10^{-1}$} &
\textcolor{red}{$0.7^{+0.1}_{-0.1}$} &
\textcolor{red}{$112^{+11}_{-9}$} &
\textcolor{red}{$4.3^{+0.3}_{-0.3}$} &
\textcolor{red}{$0.7$} \\
\enddata
\end{deluxetable*}

\end{appendix}

\bibliography{citation.bib}{}
\bibliographystyle{aasjournalv7}
\end{CJK*}
\end{document}